\documentclass[preprint]{aastex}
\usepackage{emulateapj5}

\newcommand{\Msol}{M$_{\odot}$}
\newcommand{\Mbol}{M$_{bol}$}
\newcommand{\Mjup}{M$_{\mathrm{JUP}}$}

\newcommand{\Msini}{M\,$\sin i$}

\newcommand{\ms}{m\,s$^{-1}$}

\newcommand{\rhk}{R$^{\prime}_{HK}$}
\newcommand{\lrhk}{$\log$\,R$^{\prime}_{HK}$}
\newcommand{\au}{a.u.}
\received{July 2002}
\accepted{}
\journalid{}{}
\slugcomment{Submitted to  {\em The Astrophysical Journal}.}

\shortauthors{Tinney et al.}
\shorttitle{Four new planets}

\begin{document}
\title{Four new planets  orbiting somewhat metal-enriched stars\altaffilmark{1} }

\author{C.G. Tinney\altaffilmark{2}, R. Paul Butler\altaffilmark{3}, 
        Geoffrey W. Marcy\altaffilmark{4,5}, Hugh R.A. Jones\altaffilmark{6}, 
        Alan J. Penny\altaffilmark{7}, Chris McCarthy\altaffilmark{3}, 
        Brad D. Carter\altaffilmark{8}, Jade Bond\altaffilmark{9,2}}

\altaffiltext{1}{Based on observations obtained at the
    Anglo--Australian Telescope, Siding Spring, Australia.}
\altaffiltext{2}{Anglo-Australian Observatory, PO Box 296, Epping. 1710. 
Australia. {\tt cgt@aaoepp.aao.gov.au}}
\altaffiltext{3}{Carnegie Institution of Washington,Department of Terrestrial Magnetism,
       5241 Broad Branch Rd NW, Washington, DC 20015-1305}
\altaffiltext{4}{Department of Astronomy, University of California, Berkeley, CA, 94720}
\altaffiltext{5}{Department of Physics and Astronomy, San Francisco State University, San Francisco, CA 94132.}
\altaffiltext{6}{Astrophysics Research Institute, Liverpool John Moores University, 
       Twelve Quays House, Egerton Wharf, Birkenhead CH41 1LD, UK}
\altaffiltext{7}{Rutherford Appleton Laboratory, Chilton, Didcot, Oxon OX11 0QX, U.K.}
\altaffiltext{8}{Faculty of Sciences, University of Southern Queensland, Toowoomba, 4350. Australia.}
\altaffiltext{9}{School of Physics, University of Sydney, Sydney, 2000. Australia.}

\begin{abstract}
We report the detection of four new extra-solar planets
from the Anglo-Australian Planet Search orbiting the somewhat metal-enriched
stars HD\,73526, HD\,76700, HD\,30177 and HD\,2039. The planetary companion of
HD\,76700 has a circular orbit with a period of 3.98\,d. With
M\,$\sin i$=0.197$\pm$0.017\Mjup, or 0.69 times the mass of Saturn, is
one of the lowest minimum mass extra-solar planets yet detected.
The remaining planets all have elliptical orbits with periods
ranging from 190.5\,d to 4.4\,yr.
All four planets have been found orbiting stars from a sub-sample
of twenty metal-enriched and faint (V$<$9) stars, which was added to the Anglo-Australian Planet
Search's magnitude-limited V$<$7.5 main sample in October 1998. These stars
were selected to be metal-enriched on the basis of their Str\"omgren photometry,
and their enrichment has been subsequently confirmed by detailed spectroscopic analysis.
\end{abstract}


\section{Introduction}

The Anglo-Australian Planet Search (AAPS) is a long-term planet
detection program which aims to perform extra-solar
planet detection and measurement at the highest possible precision.
Together with programmes using similar techniques on the
Lick 3\,m and Keck I 10\,m telescopes \citep{fischer01,vmba00}, 
it provides all-sky planet search coverage for inactive F,G,K and M dwarfs
down to a magnitude limit of V=7.5. Initial results from
this programme \citep{AAPSI,AAPSII,AAPSIII,AAPSIV,AAPSV,AAPSVI} demonstrate  that it 
achieves long-term,
systematic velocity precisions of 3\,\ms\ or better, for suitably stable stars,
down to our main sample magnitude limit of V$<$7.5.

AAPS is being carried out on the 3.9\,m Anglo-Australian
Telescope (AAT), using the University College London Echelle Spectrograph (UCLES)
and an I$_2$ absorption cell.
UCLES is operated in its 31\,lines\,mm$^{-1}$ mode. Prior to 2001 September,
it was used with  a MIT/LL 2048$\times$4096 15$\mu$m pixel CCD, and since then has
been used with an EEV 2048$\times$4096 13.5$\mu$m pixel CCD. 
Our target sample includes 178 FGK stars with $\delta < -20$\arcdeg\ and V$<$7.5. 
Where age/activity information is available 
from \rhk\ indices -- see for example \citet{hsdb96,CaHKI} -- we require
target stars to have $\log$\rhk $>$ -4.5 (corresponding to ages
greater than 3\,Gyr). The observing and data processing 
procedures follow that described in \citet{bmwmd96} and  
\citet{AAPSII}.

In addition to our primary sample of V$<$7.5 dwarfs, a small sub-sample
of twenty fainter dwarfs (down to V$<$9)  was added in October 1999, 
following suggestions that metal-enriched stars seemed to be preferentially
revealing planets (see eg. \citet{laughlin00} and references therein). 
These stars all had 
$uvby$ photometry suggesting metal-enrichment.
All stars in this fainter sub-sample were observed
with a maximum exposure time of 300s regardless of observing conditions.
As a result, the velocity precisions achieved for these targets are not as high
as those demonstrated for the AAPS main sample \citep{AAPSIV,AAPSII}.

\section{Characteristics of the Host Stars}

{\bf HD\,30177} (HIP\,21850, SAO\,233633) is a chromospherically inactive 
({\lrhk}=$-5.08$; \citet{CaHKI}) G8V star \citep{mssI}. 
HD\,30177 was observed 105 times by the HIPPARCOS satellite
and found to be photometrically stable 
with a standard deviation of 0.017 magnitudes.
Its HIPPARCOS parallax of 18.3$\pm$0.8\,mas implies absolute magnitudes
of M$_V$=4.72$\pm$0.09 \citep{esa97}
and \Mbol=4.36$\pm$0.10 \citep{allenIV} .
%
%
Str\"omgren $uvby$ photometry from the General Catalogue
of Photometric Data\footnote{Data obtained directly from 
{\tt http://obswww.unige.ch/gcpd/gcpd.html}} \citep{hm97}, 
together with the $uvby$ calibrations of
\citep{sn89} indicate a metallicity of [Fe/H]=$+0.20{\pm}0.16$. 
A detailed metallicity analysis has been performed for all the
G-dwarfs in the AAPS \citep{Bond02}, which indicates a metallicity
of [Fe/H]=$+0.19{\pm}0.09$, in agreement with the $uvby$-based estimate. The
$b-y$ colour of HD\,30177 indicates T$_{\mathrm eff}$=5320$\pm$20K \citep{hm97,olsen84}.
From HD30177's {\lrhk}=$-5.08$ we estimate intrinsic stellar velocity jitter
to be in the range 2-4\,\ms\ \citep{saar98}.
\placefigure{models} 

Figure \ref{models} shows evolutionary tracks at [Fe/H]=+0.2 and +0.1
for stars of near-Solar mass from the compilation of \citet{girardi00}.
The [Fe/H]=+0.1 tracks were obtained by interpolation between the
[Fe/H]=+0.2 and [Fe/H]=0.0 tracks in \citet{girardi00}.
Based on these models and the metallicity measurements for 
HD\,30177, its mass is estimated to be 0.95$\pm$0.05\Msol,
and it would appear (in common with HD\,73526 and HD\,76700)
to be beginning its evolution off the main sequence.

%
%

{\bf HD\,73526} (HIP\,42282, SAO\,220191) is a G6V dwarf \citep{mssII} for which
no \rhk\ estimate is currently available. It was observed 137 times by HIPPARCOS,
but its relative faintness means that only a 0.02 magnitude standard deviation
limit is placed on its photometric stability.
Its HIPPARCOS parallax is 10.6$\pm$1.0\,mas,
which indicates M$_V$=4.1$\pm$0.2 and \Mbol=3.7$\pm$0.2 \citep{esa97,allenIV}. 
As for HD\,30177 (see above), Str\"omgren $uvby$ photometry indicates
a metallicity of [Fe/H]=+0.10$\pm$0.16 \citep{hm97,sn89}, while a
detailed metallicity analysis gives [Fe/H]=+0.11$\pm$0.10 \citep{Bond02}.
The GCPD
$b-y$ colour of HD\,73526 indicates T$_{\mathrm eff}$=5450$\pm$20K \citep{hm97,olsen84}.
Based on the evolutionary tracks shown in Figure \ref{models}, 
HD\,73526 is estimated to have a mass of 1.05$\pm$0.05\Msol.

%
%

\begin{figure}
\epsscale{0.3}
\plotone{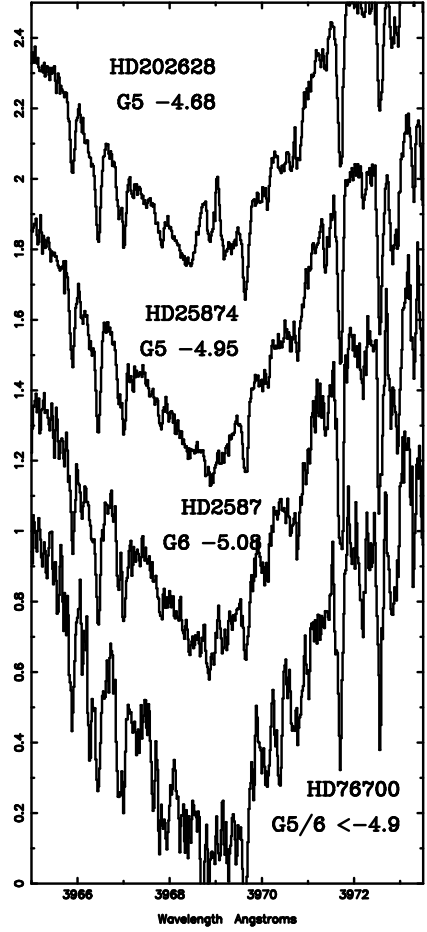}
\caption{UCLES spectrum in the region of the Ca~H line for HD\,76700 and
several comparison objects with \lrhk measurements from \citet{CaHKI}.
For each object the spectral type and measured \lrhk value is shown.
Based on these we assign an upper limit to the \lrhk value for HD\,76700
of -4.9.
}
\label{cahk}
\end{figure}

{\bf HD\,76700} (HIP\,43686, SAO\,250370, LTT\,3291) is catalogued as a G6V dwarf by SIMBAD,
as G5 by the Henry-Draper catalogue \citep{hd} and as G8V by HIPPARCOS \citep{esa97}.
The HIPPARCOS B-V colour ($+$0.745) would indicate a G8 spectral type is more likely than
G5 or G6, though the GCPD $b-y$ colour \citep{hm97} would indicate a G6 type is
more appropriate.
No \rhk\ estimate is available. It was observed 109 times by HIPPARCOS,
which found it to have constant magnitude with a standard deviation of 0.012 magnitude.
Its HIPPARCOS parallax is 16.8$\pm$0.7\,mas,
which indicates M$_V$=4.3$\pm$0.1 and \Mbol=3.9$\pm$0.2 \citep{esa97,lang}. 
Str\"omgren $uvby$ photometry indicates
a metallicity of [Fe/H]=+0.14$\pm$0.16 \citep{hm97,sn89}, while a
detailed metallicity analysis gives [Fe/H]=+0.10$\pm$0.11 \citep{Bond02}.
The
GCPD $b-y$ colour indicates a T$_{\mathrm eff}$=5423$\pm$20K \citep{hm97,olsen84}.
Based on the evolutionary tracks shown in Figure \ref{models}, 
HD\,76700 is estimated to have a mass of 1.00$\pm$0.05\Msol.
There is no published \lrhk measurement for HD\,76700. Because
the velocity amplitude observed in this star is small (26$\pm$2\,\ms),
a Ca\,HK spectrum was acquired to determine the level at which our results
could be affected by velocity jitter. Figure \ref{cahk} shows this spectrum for HD\,76700
compared to those for several objects of similar spectral type, 
along with their spectral types and \lrhk values \citep{CaHKI}.
HD\,76700 shows no evidence for a line reversal.
Based on this comparison we assign an upper limit to the \lrhk for
HD\,76700 of -4.9, from which we estimate its intrinsic stellar velocity jitter
to be in the range 3-6\,\ms\ or less \citep{saar98}.

%
%
%
%

{\bf HD\,2039} (HIP\,1931, SAO\,23205) is a chromospherically inactive 
({\rhk}=$-4.91$; \citet{CaHKI}) star classified as G2/G3 IV/V by \citet{mssI}. 
It was observed 147 times by HIPPARCOS,
but its relative faintness means that only a 0.022 magnitude standard deviation
limit is placed on its photometric stability.
Its HIPPARCOS parallax is 11.1$\pm$1.1\,mas,
which indicates M$_V$=4.25$\pm$0.24 and \Mbol=4.22$\pm$0.25 \citep{esa97,allenIV}. 
There are multiple measures of this star's Str\"omgren $uvby$ photometry, the
mean of which indicate a metallicity of [Fe/H]=+0.10$\pm$0.16 \citep{hm97,sn89}.
A detailed metallicity analysis gives [Fe/H]=+0.10$\pm$0.11 \citep{Bond02}.
The
$b-y$ colour of HD\,2039 indicates T$_{\mathrm eff}$=5675$\pm$20K \citep{hm97,olsen84}.
Based on the evolutionary tracks shown in Figure \ref{models}, 
HD\,2039 is estimated to have a mass of 0.98$\pm$0.05\Msol.

%
%
%
%
%

\section{Radial Velocity Observations and Orbital Solutions}
\label{velocities}

\begin{figure}
\epsscale{0.5}
\plotone{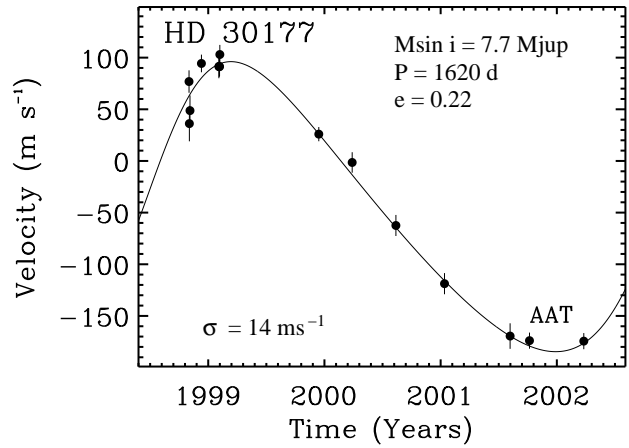}
\caption{AAT Doppler velocities for HD\,30177 from 1998 November to
2002 May.   The solid line is a best
fit Keplerian with the parameters shown
in Table \ref{orbits1}.The rms of the velocities about the fit is 14\,\ms. 
Assuming 0.95\,\Msol\ for the primary,
the minimum (\Msini) mass of the companion is 7.7$\pm$1.5\,\Mjup, and
the semi-major axis is 2.6$\pm$0.9\,a.u.}
\label{hd30177_rv_curve}
\end{figure}

Fifteen observations of HD\,30177 are listed in Table \ref{vel30177}. 
The 
column labelled ``Uncertainty'' is the velocity uncertainty produced by our
least-squares fitting procedure. This fit simultaneously determines the Doppler shift and
the spectrograph point-spread function (PSF) for each observation
made though the iodine cell, given an iodine absorption spectrum and
an ``iodine free'' template spectrum of the object \citep{bmwmd96}. 
The uncertainty is derived for each measurement by taking the mean of four hundred useful
spectral regions (each 2\,\AA\ long) from each exposure.
This uncertainty includes the effects of photon-counting uncertainties,
residual errors in the spectrograph PSF model, and variation in the underlying 
spectrum between the template and ``iodine'' epochs. All velocities are measured relative to the 
zero-point defined by the template observation. 
Only observations
where the uncertainty is less than twice the median uncertainty are
listed.
These data are shown in Figure \ref{hd30177_rv_curve}. 
The figure shows the best-fit Keplerian model for the
data, with the resultant orbital parameters listed in Table \ref{orbits1}. 
Due to HD\,30177's faintness compared to the AAPS main sample,
the residuals about the fit (14\ms) are significantly higher than
the baseline 3\ms\ precision level demonstrated for the
main sample \citep{AAPSII,AAPSIV}.

\placetable{vel30177}  

\placefigure{hd30177_rv_curve} 

\placetable{orbits1}  

\begin{figure}
\epsscale{0.5}
\plotone{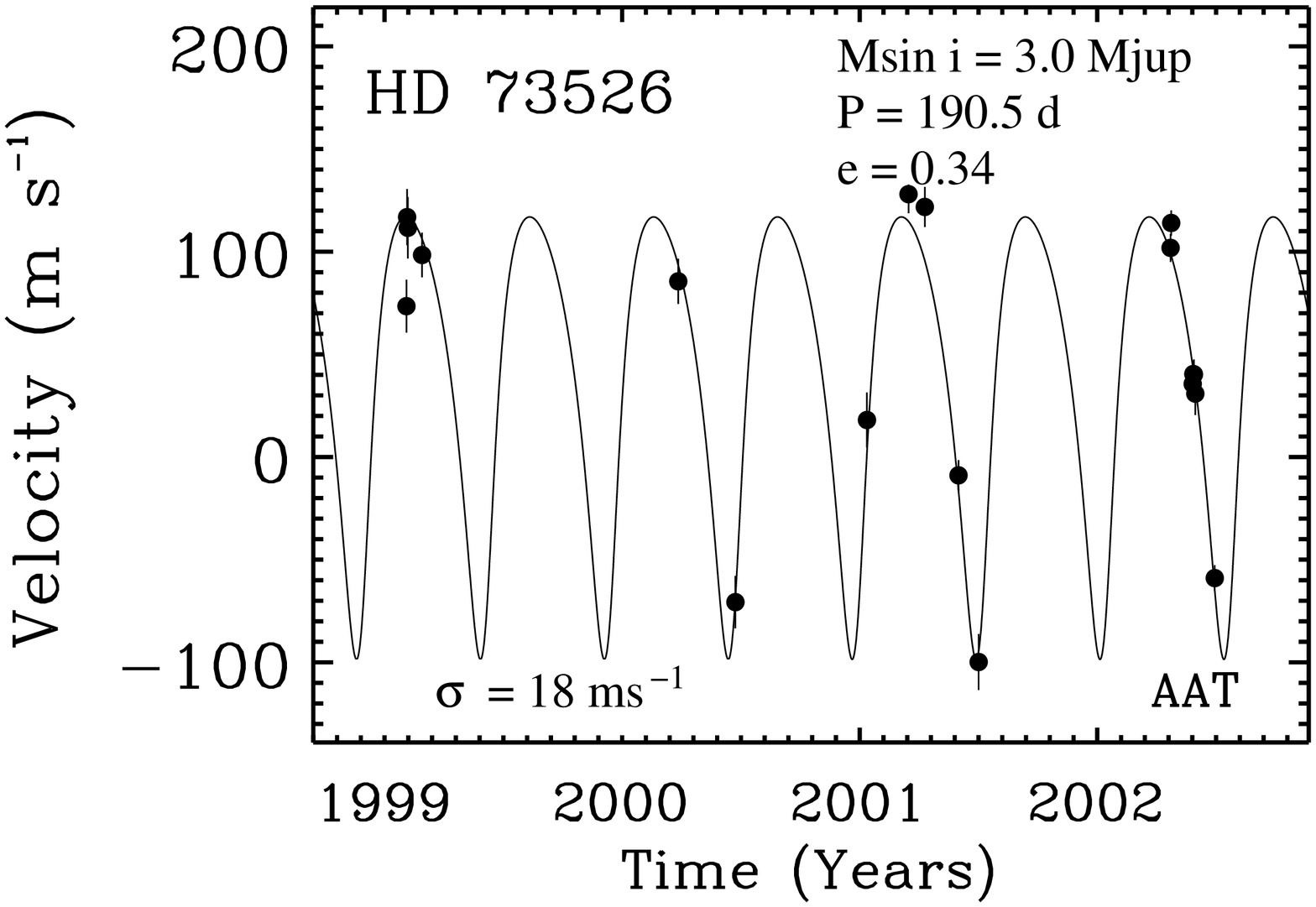}
\caption{AAT Doppler velocities for HD\,73526 from 1999 January to
2002 June. The solid line is a best fit Keplerian with the parameters shown
in Table \ref{orbits1}.
The rms of the velocities about the fit is 18\,\ms. Assuming 1.05\,\Msol\ for the primary,
the minimum (\Msini) mass of the companion is 3.0$\pm$0.3\,\Mjup, and
the semi-major axis is 0.66$\pm$0.05\,a.u.}
\label{hd73526_rv_curve}
\end{figure}

\begin{figure}
\epsscale{1.0}
\plottwo{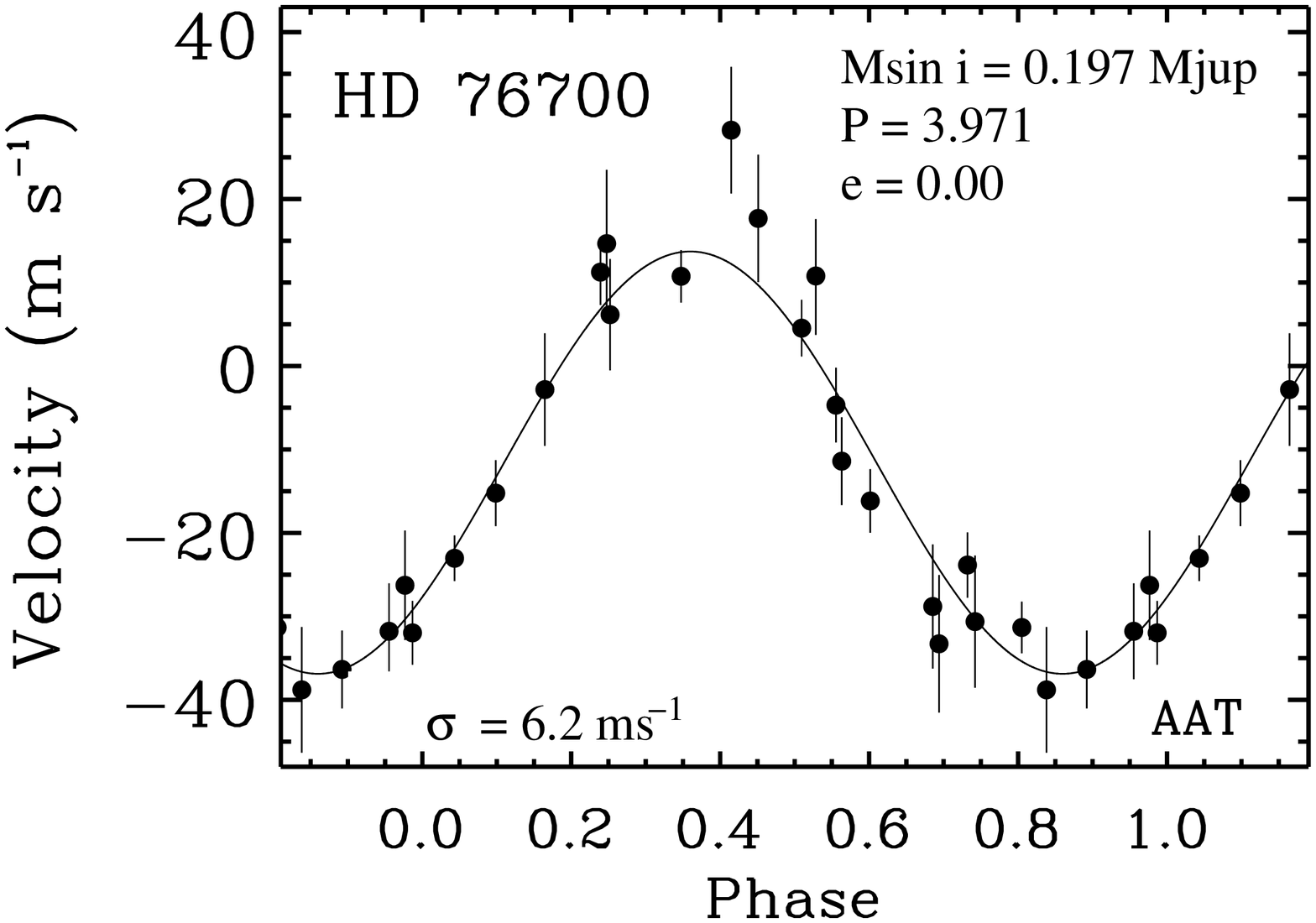}{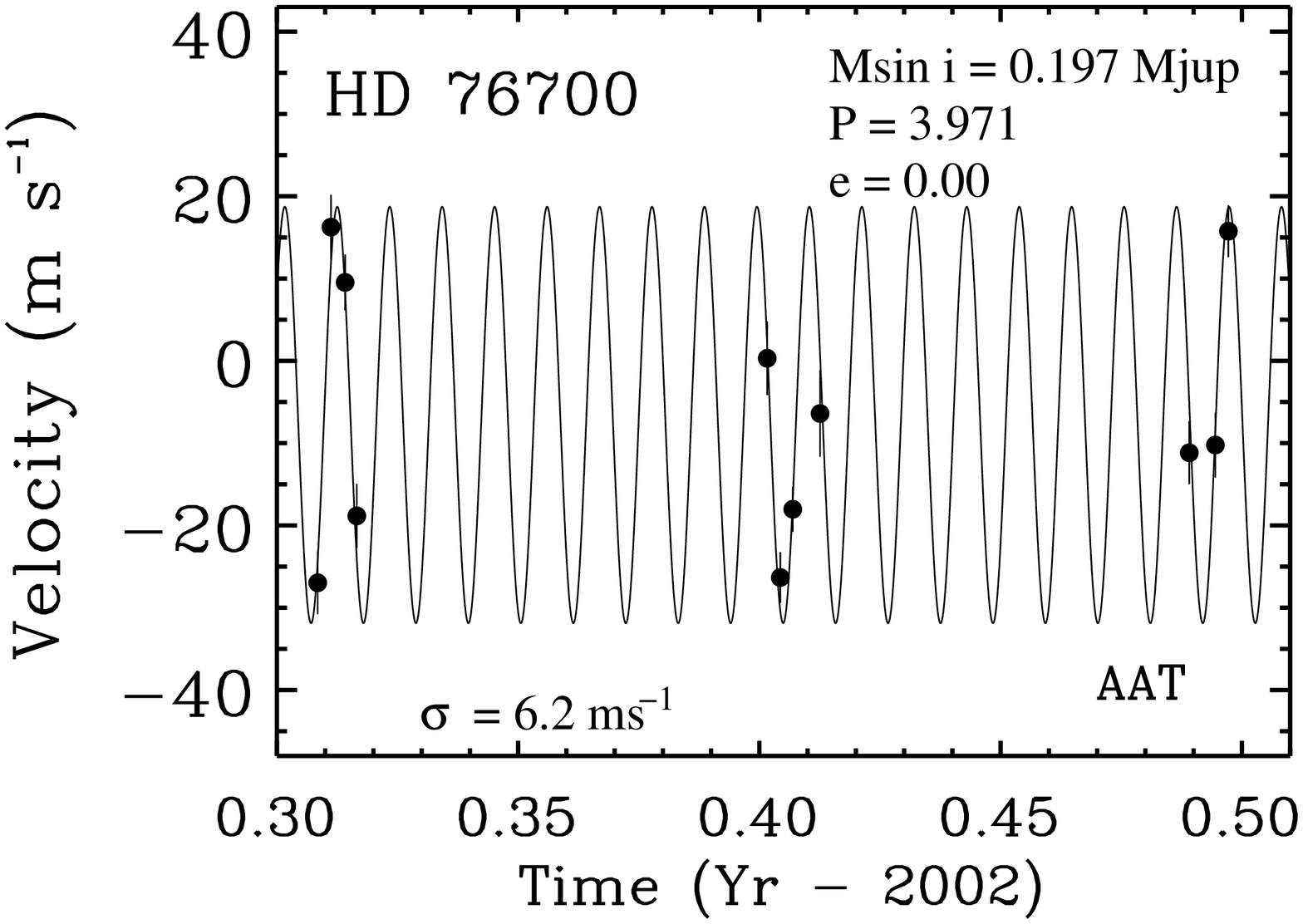}
\caption{AAT Doppler velocities for HD\,76700 from 1999 February to
2002 June, phased at the best fit Keplerian period of 3.971\,d {\em (left panel)} and
plotted unphased from the last three observing runs in 2002 {\em (right panel)}. The plotted
Keplerian has the parameters shown in Table \ref{orbits1}.
The rms of the velocities about the fit is 6.2\,\ms. Assuming 1.0\,\Msol\ for the primary,
the minimum (\Msini) mass of the companion is 0.197$\pm$0.017\,\Mjup, and
the semi-major axis is 0.049$\pm$0.004\,a.u.}
\label{hd76700_rv_curve}
\end{figure}

Eighteen observations of HD\,73526 are listed in Table \ref{vel73526}, 
and they are shown in Figure \ref{hd73526_rv_curve}
along with a Keplerian fit to the data with
the orbital parameters listed in Table \ref{orbits1}. Note again the larger
residuals (17\,\ms) compared to the AAPS main sample. 
Twenty-four observations of HD\,76700 are listed in Table \ref{vel76700}, 
and they are shown in Figure \ref{hd76700_rv_curve}
along with a Keplerian fit to the data with
the orbital parameters listed in Table \ref{orbits1}. 
The thirty-six observations of HD\,2039 are listed in Table \ref{vel2039}, 
and they are shown in Figure \ref{hd76700_rv_curve}
along with a Keplerian fit to the data with
the orbital parameters listed in Table \ref{orbits1}. 

\placetable{vel73526}  

\placefigure{hd73526_rv_curve} 

\placetable{vel76700}  

\placefigure{hd76700_rv_curve} 

\placetable{vel2039}  

\begin{figure}
\epsscale{0.5}
\plotone{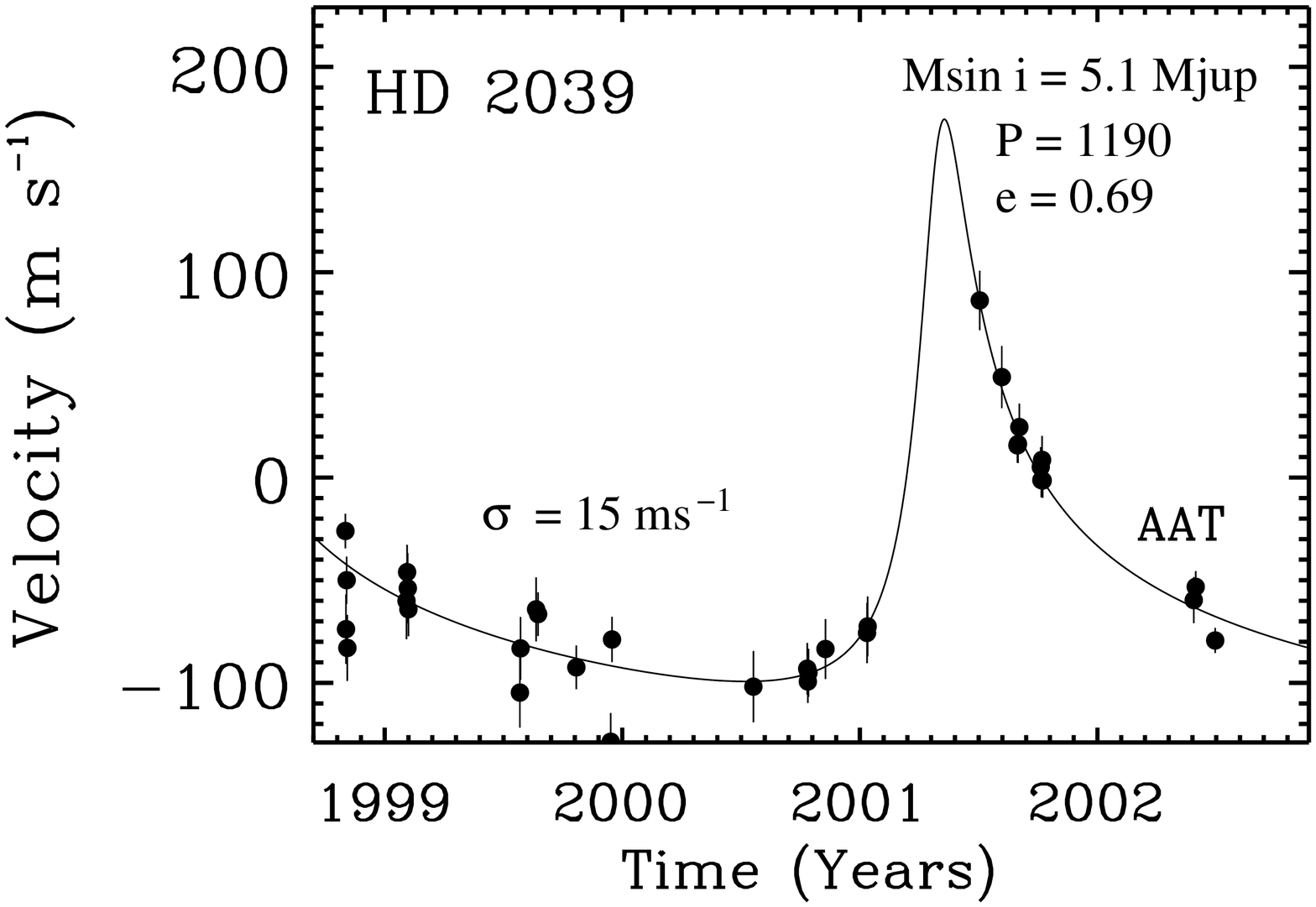}
\caption{AAT Doppler velocities for HD\,2039 from 1998 November to
2002 June, with the best fit Keplerian with 
 parameters shown in Table \ref{orbits1}.
The rms of the velocities about the fit is 15\,\ms. Assuming 0.98\,\Msol\ for the primary,
the minimum (\Msini) mass of the companion is 5.1$\pm$1.7\,\Mjup, and
the semi-major axis is 2.2$\pm$0.2\,a.u.}
\label{hd2039_rv_curve}
\end{figure}

\placefigure{hd2039_rv_curve} 

\section{Discussion}

The resultant minimum companion mass for HD\,76700 is \Msini\,=\,0.197$\pm$0.017\,\Mjup,  
with an orbital semi-major axis $a$\,=\,0.049$\pm$0.004\,au and eccentricity $e$=0.00$\pm$0.04.
This zero eccentricity is consistent with the expectation that a planet
with a period of just 3.971$\pm$0.001\,d will almost certainly lie in an orbit which 
has been tidally circularised \citep{mb98}. The resulting orbital parameters for HD\,76700
place it amongst the planetary companions with the lowest known minimum masses. HD\,76700,
joins HD\,49674, HD\,16141, HD\,168746 \& HD\,46375 \citep{butler02b, mbv00, pepe02} in the
group of extra-solar planetary companions with measured minimum masses less than a 
Saturn mass (0.299\,\Mjup).

The remaining three extra-solar planets all have elliptical orbits, though the
ellipticity of the orbit for the companion to HD\,30177 ($e$=0.22$\pm$0.17)
is not different from zero with great significance. 
As further data are acquired over the coming years this parameter will become 
far better constrained,
and it is possible that this extra-solar planet could turn out to be in a substantially
circular orbit. If so this system would join with the other known nearly circular systems with
gas-giant planets lying in orbits between where the Earth and Jupiter lie in our
own Solar System ($\epsilon$\,Ret, HD\,4208, the outer components of 47\,UMa, HD\,28185 
\citep{AAPSIV}). These extra-solar planets would seem to indicate that gas-giants 
exist in nearly circular orbits with semi-major axes all the way out to, and beyond
that of Jupiter, as confirmed recently by the detection of the outer planet
in the 55\,Cnc system \citep{55Cnc}.

\begin{figure}
\epsscale{1.0}
\plottwo{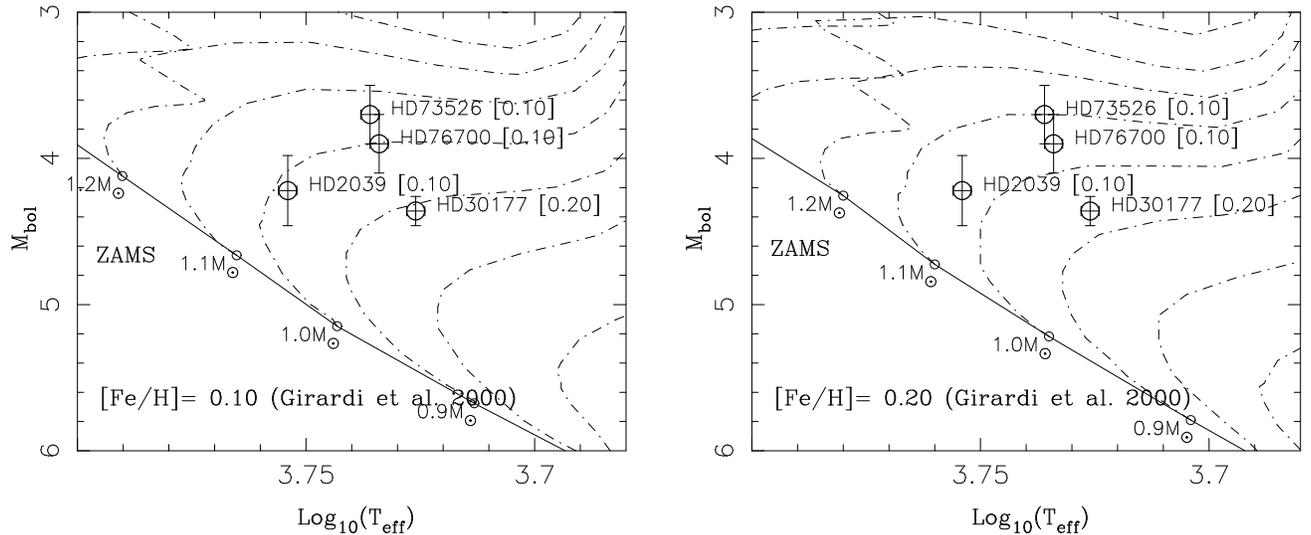}{f2.eps}
\caption{Evolutionary tracks from the compilation of \citet{girardi00}. The Zero Age Main
Sequence (ZAMS) is shown as a solid line, with evolutionary models at the
indicated masses shown as {\em dot-dashed} lines. The open circles and uncertainties
indicate the stars under discussion in this paper, with their estimated uncertainties indicated
in brackets. The [Fe/H]=+0.2 tracks 
are directly from \citet{girardi00}, while the [Fe/H]=+0.1 tracks were obtained by interpolation
between the catalogued [Fe/H]=+0.2 and 0.0 models.}
\label{models}
\end{figure}

Of the twenty metal-enriched stars included as a sub-sample along with
our main sample in late 1999, five have revealed the presence of planetary companions (the four
planets discussed here -- HD\,30177, 73526, 76700 and 2039 -- along with the previously
known companion to HD\,83443 \citep{AAPSV,mayor_83443}). This gives us a lower limit 
(there may be longer period or lower mass planets present which we cannot yet
detect) to the
discovery rate of 25$\pm$11\% for this ``metallicity-biassed'' sub-sample. This compares with
the overall discovery rates estimated for the Keck, Lick and AAPS of $\sim$8\% (ie 8\% of
stars surveyed have planets in orbits within 3.5\,\au\ of their host stars \citep{properties})
-- a difference which, while not of great statistical significance, is not unexpected given
that extra-solar planets seem to be being found preferentially around metal-enriched stars
(eg. \citet{reid02,laughlin00} and references therein). 
It is also interesting to note (even if perhaps not statistically significant) that
all four of the stars in this paper would seem to be beginning their evolution off
the main sequence (see Fig \ref{models}).
 
These results would suggest
that the biassing of planet surveys toward metal-enriched host stars {\em may} offer a
benefit in the planet detection rate. However, such an increased discovery rate
must be balanced against the fact that it will produce an inherently biassed
sample of extra-solar planets. With the total number of extra-solar planets still
numbering less than a hundred, and the parameters of this ensemble of planets
still poorly placed in a scheme of extra-solar planetary formation and evolution, now is not
the time for planet searches to begin biassing their large surveys in the
chase for better ``hit rates'' at the expense of scientific utility.

\acknowledgments
The Anglo-Australian Planet Search team would like to gratefully acknowledge the support
of Dr Brian Boyle, Director of the AAO,  the contributions and
assistance of Drs Stuart Ryder and Debra Fischer,
and the superb technical support which has been
received throughout the programme from AAT staff - in particular R.Patterson, 
D.Stafford, F.Freeman, S.Lee, J.Pogson, G.Kitley and J. Stevenson. We further acknowledge support 
by; the partners of the Anglo-Australian Telescope Agreement (CGT,HRAJ,AJP);
    NASA grant NAG5-8299 \& NSF grant AST95-20443 (GWM);
    NSF grant AST-9988087 (RPB); and Sun Microsystems.
NSO/Kitt Peak FTS data used here were produced by NSF/NOAO.

\begin{deluxetable}{lrr}
\tablenum{1}
\tablecaption{Velocities for HD\,30177}
\tablewidth{0pt}
\tablehead{
JD$^a$ & RV$^a$ & Uncertainty \\
(-2451000)   &  (\ms) & (\ms)
}
\startdata
   118.0974  & 77.9  & 10.9 \\
   119.1924  & 37.2  & 17.1 \\
   121.1514  & 49.7  & 14.8 \\
   157.1022  & 95.4  & 8.6 \\
   211.9834  & 92.4  & 11.3 \\
   212.9660  & 92.5  & 10.4 \\
   213.9995  & 92.3  & 9.1 \\
   214.9506  & 104.0  & 9.3 \\
   525.9973  & 26.9  & 6.9 \\
   630.9156  & -0.5  & 9.9 \\
   768.3296  & -61.5  & 10.2 \\
   921.1075  & -117.8  & 10.2 \\
  1127.3205  & -168.5  & 12.4 \\
  1188.2532  & -172.9  & 7.8 \\
  1358.9181  & -173.4  & 7.8 \\
\enddata
\tablenotetext{a}{Julian Dates (JD) are barycentric. Radial Velocities
(RV) are barycentric, but have an arbitrary zero-point determined by
the radial velocity of the template, as described in Section~\ref{velocities}}
\label{vel30177}
\end{deluxetable}

\clearpage

%
%
%

\begin{deluxetable}{lcccc}
\tablenum{2}
\tablecaption{Orbital Parameters}
\tablewidth{0pt}
\tablehead{
\colhead{Parameter}            & \colhead{HD\,30177}
                                                   & \colhead{HD\,73526} & \colhead{HD\,76700} & \colhead{HD\,2039}
                                                   }
\startdata
Orbital period $P$ (d)          &   1620$\pm$800     &  190.5$\pm$3.0    &  3.971$\pm$0.001   &   1190$\pm$150               \\
Velocity amp. $K$ (\ms)         &    140$\pm$10      &  108$\pm$8        &  25$\pm$2           &   136$\pm$30                  \\
Eccentricity $e$                &  0.22$\pm$0.17     & 0.34$\pm$0.08     &  0.00$\pm$0.04      &   0.69$\pm$0.15              \\
$\omega$ (\arcdeg)              &    288$\pm$40      & 207$\pm$30        &  -                  &   333$\pm$20                 \\
$a_1 \sin i$ (km)               &  (3.0$\pm$1.5)$\times$10$^6$                                 
                                                     & (0.265$\pm$0.01)$\times$10$^6$
                                                                         &  1381$\pm$1.2         & (1.61$\pm$0.45)$\times$10$^6$  \\                                                                                               
Periastron Time (JD-245000)     & 1027$\pm$200       &  951$\pm$12       &  1212.9$\pm$0.1  &  836$\pm$150                 \\
\Msini\ (\Mjup)                 &  7.7$\pm$1.5       &  3.0$\pm$0.3      &  0.197$\pm$0.017      &  5.1$\pm$1.7                 \\
a (AU)                          &  2.6$\pm$0.9       &  0.66$\pm$0.05    &  0.049$\pm$0.004    &  2.2$\pm$0.2                 \\
RMS about fit (\ms)             &  14                &  18               &  6.2                &  15                          \\
\enddata
\label{orbits1}
\end{deluxetable}

%
%
%
%

\begin{deluxetable}{lrr}
\tablenum{3}
\tablecaption{Velocities for HD\,73526}
\tablewidth{0pt}
\tablehead{
JD$^a$ & RV$^a$ & Uncertainty \\
(-2451000)   &  (\ms) & (\ms)
}
\startdata
   212.1302  & 23.5  & 12.9 \\
   213.1314  & 66.8  & 13.7 \\
   214.2389  & 61.6  & 14.9 \\
   236.1465  & 48.4  & 10.9 \\
   630.0280  & 35.5  & 11.0 \\
   717.9000  & -120.7  & 12.8 \\
   920.1419  & -32.0  & 13.4 \\
   984.0378  & 78.0  & 9.3 \\
  1009.0976  & 71.8  & 9.7 \\
  1060.8844  & -58.9  & 7.4 \\
  1091.8465  & -149.8  & 13.6 \\
  1386.9003  & 51.8  & 6.8 \\
  1387.8921  & 64.0  & 6.2 \\
  1420.9248  & -14.4  & 7.8 \\
  1421.9199  & -9.6  & 6.5 \\
  1422.8602  & -9.8  & 7.3 \\
  1424.9237  & -19.1  & 10.4 \\
  1454.8529  & -109.0  & 6.3 \\
\enddata
\tablenotetext{a}{As for Table \ref{vel30177}}
\label{vel73526}
\end{deluxetable}

%
%
%
%

\begin{deluxetable}{lrr}
\tablenum{4}
\tablecaption{Velocities for HD\,76700}
\tablewidth{0pt}
\tablehead{
JD$^a$ & RV$^a$ & Uncertainty \\
(-2451000)   &  (\ms) & (\ms)
}
\startdata
   212.1565  & 9.2  & 6.7 \\
   213.1501  & 40.3  & 7.6 \\
   214.2583  & -21.3  & 8.2 \\
   274.0177  & -18.6  & 7.9 \\
   530.1791  & 26.7  & 8.8 \\
   683.8938  & -19.8  & 5.7 \\
   920.1606  & 29.7  & 7.6 \\
   984.0068  & 22.8  & 7.1 \\
  1009.0638  & -26.8  & 7.5 \\
  1060.9036  & -24.3  & 4.7 \\
  1091.8517  & -16.8  & 7.4 \\
  1129.8425  & 18.2  & 6.7 \\
  1359.0760  & -14.3  & 6.6 \\
  1386.9145  & -20.0  & 3.8 \\
  1387.9170  & 23.2  & 3.9 \\
  1388.9898  & 16.5  & 3.4 \\
  1389.8741  & -11.8  & 3.9 \\
  1420.9418  & 7.3  & 4.5 \\
  1421.9331  & -19.3  & 3.1 \\
  1422.8793  & -11.0  & 2.7 \\
  1424.9439  & 0.6  & 5.3 \\
  1452.8943  & -4.2  & 3.8 \\
  1454.8682  & -3.2  & 3.9 \\
  1455.8565  & 22.8  & 3.1 \\
\enddata
\tablenotetext{a}{As for Table \ref{vel30177}}
\label{vel76700}
\end{deluxetable}

%
%

%

\begin{deluxetable}{lrr}
\tablenum{5}
\tablecaption{Velocities for HD\,2039}
\tablewidth{0pt}
\tablehead{
JD$^a$ & RV$^a$ & Uncertainty \\
(-2451000)   &  (\ms) & (\ms)
}
\startdata
   118.0578  & 24.0  & 8.4 \\
   118.9610  & -23.8  & 16.9 \\
   119.9445  & 0.0  & 11.5 \\
   121.0385  & -33.0  & 16.1 \\
   211.9514  & -10.1  & 18.7 \\
   212.9234  & 3.9  & 13.3 \\
   213.9749  & -3.9  & 17.0 \\
   214.9171  & -14.1  & 13.2 \\
   386.3227  & -54.7  & 17.1 \\
   387.2981  & -33.2  & 15.3 \\
   411.2293  & -14.2  & 15.6 \\
   414.2585  & -16.5  & 10.6 \\
   473.0883  & -42.5  & 10.6 \\
   525.9286  & -78.7  & 13.9 \\
   527.9226  & -28.9  & 11.0 \\
   745.2702  & -51.8  & 17.4 \\
   828.0703  & -43.1  & 12.5 \\
   828.9944  & -49.3  & 10.4 \\
   829.9757  & -45.1  & 11.7 \\
   856.0702  & -33.5  & 14.6 \\
   919.9434  & -25.7  & 14.7 \\
   920.9672  & -22.5  & 14.5 \\
  1093.2947  & 136.2  & 14.4 \\
  1127.2341  & 98.9  & 15.1 \\
  1151.2230  & 65.7  & 8.4 \\
  1152.0860  & 66.4  & 9.2 \\
  1154.2124  & 74.6  & 11.4 \\
  1187.0957  & 55.1  & 9.7 \\
  1188.0300  & 48.7  & 8.4 \\
  1189.1502  & 58.6  & 11.6 \\
  1190.0932  & 48.5  & 8.4 \\
  1422.3281  & -9.7  & 11.2 \\
  1425.3322  & -3.2  & 7.7 \\
  1455.2853  & -29.4  & 6.1 \\
\enddata
\tablenotetext{a}{As for Table \ref{vel30177}}
\label{vel2039}
\end{deluxetable}

\clearpage

\end{document}